\documentstyle[epsf]{ptptex}

\title{%
  Renormalizable Abelian-Projected Effective Gauge Theory\\
  Derived from Quantum Chromodynamics%
  }

\author{%
  Kei-Ichi K{\sc ondo}$^{1,2,}$%
    \footnote{E-mail: {\tt kondo@cuphd.nd.chiba-u.ac.jp}}
  and Toru S{\sc hinohara}$^{2,}$%
    \footnote{E-mail: {\tt sinohara@cuphd.nd.chiba-u.ac.jp}}
  }

\inst{%
  ${}^1$Department of Physics, Faculty of Science, Chiba University,
         Chiba 263-8522 Japan\\
  ${}^2$Graduate School of Science and Technology, Chiba University,
         Chiba 263-8522 Japan
  }

\abst{%
We show that an effective Abelian gauge theory can be obtained as a renormalizable theory from QCD in the maximal Abelian gauge.
The derivation improves in a systematic manner the previous version that was obtained by one of the authors and was referred to as the Abelian-projected effective gauge theory.
This result supports the view that we can construct an effective Abelian gauge theory from QCD without losing characteristic features of the original non-Abelian gauge theory.  In fact, it is shown that the effective coupling constant in the resulting renormalizable theory has a renormalization-scale dependence governed by the $\beta$-function that is  exactly the same as that of the original Yang-Mills theory, irrespective of the choice of gauge fixing parameters of the maximal Abelian gauge and the parameters used for identifying the dual variables.
Moreover, we evaluate the anomalous dimensions of the fields and parameters in the resultant theory.
By choosing the renormalized parameters appropriately, we can switch the theory into an electric or a magnetic theory.
}
\recdate{November 3, 2000}

\notypesetlogo

\begin{document}

\maketitle


\section{Introduction} %

It is widely believed that quantum chromodynamics (QCD) as a non-Abelian gauge theory with color gauge group $SU(3)$ can describe quark and gluon confinement and more generally color confinement, although there is no rigorous proof to this time.  
Various physical phenomena that we can observe in experiments should be gauge independent.  However, the quantized gauge field theory is consistently formulated only in gauge-fixed form, at least in the continuum formulation, in contrast to the lattice formulation.  Even QCD is not an exception.
Therefore, the continuum QCD with a gauge-fixing term and the associated Faddeev-Popov (FP) ghost term can be written only in a specific (although arbitrary) gauge.
Traditionally, the Lorentz gauge 
$\partial_\mu {\cal A}^\mu = 0$ has been extensively examined as a manifest covariant gauge, perhaps because quantum electrodynamics (QED) was successfully formulated in this gauge.
Therefore there are numerous works on the Lorentz gauge.  However, it turns out that the Lorentz gauge in the non-Abelian gauge theory is plagued by Gribov ambiguities.  It is not clear how the color confinement criteria in the Lorentz gauge are compatible with intuitive pictures of quark confinement represented by the dual superconductor picture \cite{dsc} and QCD strings.
\par
The developments in the understanding of quark confinement in the last decade \cite{review} indicate that, within  the present technology, the maximal Abelian (MA) gauge \cite{KLSW87} realizing the idea of the Abelian projection \cite{tHooft81} seems to be the most promising gauge in which the dual superconductor picture of QCD vacuum could be derived directly from QCD.
Abelian dominance in the low-energy region of QCD has been confirmed.
This was conjectured in Ref.~\citen{EI82} and its discovery,
based on lattice simulations, was reported in Ref.~\citen{SY90}.
Moreover, the MA gauge is expected to be free from the Gribov problem,
due to its nonlinearity.
\par
In a previous paper,\cite{KondoI} we derived an  effective Abelian gauge
theory directly from QCD in the  MA gauge by integrating out all the
off-diagonal fields.
The resulting theory is called the Abelian-projected effective gauge theory
(APEGT).
This derivation supports the view that one can construct an effective Abelian gauge theory from QCD without losing characteristic features of the original non-Abelian gauge theory.
In fact, it was shown \cite{QR98,KondoI} that the effective coupling constant $g(\mu)$ in the APEGT has a renormalization scale $\mu$ dependence governed by the $\beta$-function $\beta(g)=\mu{dg(\mu)\over d\mu}$ that is exactly the same as that of the original Yang-Mills theory, exhibiting asymptotic freedom.
These results suggest in the analytical approach the Abelian dominance in the low-energy region of QCD.
Investigations along this direction have been carried out by one of the authors and his collaborators.\cite{KondoII,KondoIII,KondoIV,KondoV,KondoVI,KT99,KS00a}
\par
In the derivation of the APEGT,\cite{KondoI} we have introduced an auxiliary anti-symmetric tensor field $B_{\mu\nu}$ to convert the quartic off-diagonal gluon interaction into a quadratic form in the off-diagonal gluon fields.
This procedure enabled us to perform the integration over the off-diagonal gluons exactly for the gauge group $G=SU(2)$.
Moreover, the Abelian tensor field $B_{\mu\nu}$ can be regarded as a Hodge dual of a composite field that consists of the electric field (gauge potential).
This viewpoint becomes quite important in obtaining the dual theory, which is expected to be the dual Ginzburg-Landau theory (see Ref.~\citen{KondoI}).  
\par
However, there was an ambiguity as to the identification of the dual field $B_{\mu\nu}$.
In fact, the resulting  APEGT can depend on how the $B_{\mu\nu}$ is identified as the dual, although two choices were considered in the previous paper.\cite{KondoI}
Moreover, the resulting APEGT does not have a renormalizable form in the sense that the integration (or radiative correction) induces new terms which are absent in the original Lagrangian.
The failure of preserving renormalizability implies the impossibility of performing a systematic evaluation based on the APEGT.
\par
In this paper, we remove this ambiguity and the  difficulty mentioned above by deriving a renormalizable APEGT.
This renormalizable APEGT has a definite meaning irrespective of manner in which $B_{\mu\nu}$ is identified.
The renormalizable APEGT presents a promising way for performing higher-order calculations, e.g., recovery of the two-loop $\beta$-function of the original Yang-Mills theory based on the APEGT (see Ref.~\citen{KS00c}).
In this sense, this paper supplements and improves the previous work.\cite{KondoI}
In this paper, it is shown that the $\beta$-function of the original Yang-Mills theory is obtained in the APEGT, irrespective of the choice of gauge-fixing parameters for the MA gauge and also of the parameters introduced for identifying the dual variables.
Moreover, we calculate the anomalous dimensions for the diagonal (Abelian) fields.  The renormalization-group functions obtained in this way determine the scaling behavior of the vertex function in the APEGT.

This paper is organized as follows.
In \S2, we introduce the MA gauge fixing in the Yang-Mills theory
and discuss how we can identify the dual Abelian variables in the APEGT,
which will be obtained in the final step.
In \S3, we explain the meaning of APEGT and its renormalizability.
In \S4, we explain the procedure of renormalization (at the one-loop level)
to obtain the renormalizable APEGT.
Then we give the Feynman rule in the framework of the renormalized
perturbation theory.
Finally, we evaluate the renormalization group $\beta$-function and the anomalous dimensions.
In the final section we give the conclusion of the paper and discuss the
issues to be clarified in the future.

\section{Lagrangian in the MA gauge} %

\subsection{The gauge invariant part} %
The Yang-Mills Lagrangian is given by
\begin{equation}
{\cal L}_{\rm YM}
  =-\frac14\left({\cal F}_{\mu\nu}^A\right)^2 ,
\end{equation}
where 
${\cal F}_{\mu\nu}^A$ is the non-Abelian field strength defined by
\begin{equation}
 {\cal F}_{\mu\nu}^A
  =\partial_\mu {\cal A}_\nu^A
   -\partial_\nu {\cal A}_\mu^A
   +gf^{ABC}{\cal A}_\mu^B{\cal A}_\nu^C .
\end{equation}
Now we consider the decomposition of the non-Abelian gauge fields into diagonal and off-diagonal components. 
In the following we distinguish the indices as follows:
\begin{equation}
\left\{
\begin{array}{l}
\mbox{$A,B,C,\cdots$ $\rightarrow$ $SU(N)$}, \cr
\mbox{$i,j,k,\cdots$ $\rightarrow$  $U(1)^{N-1}$ (diagonal)}, \cr
\mbox{$a,b,c,\cdots$ $\rightarrow$  $SU(N)/U(1)^{N-1}$ (off-diagonal)}.
\end{array}
\right .
\end{equation}
We write the decomposition of the gauge potential into the diagonal and off-diagonal components as
\begin{equation}
  {\cal A}_\mu = {\cal A}_\mu^A T^A = A_\mu^i T^i + A_\mu^a T^a .
\end{equation}
First, the Yang-Mills Lagrangian is decomposed as
\begin{equation}
{\cal L}_{\rm YM}
 ={\cal L}_{\rm YM}^{(i)}
  +{\cal L}_{\rm YM}^{(a)},
\quad
{\cal L}_{\rm YM}^{(i)}
  =-\frac14\left({\cal F}_{\mu\nu}^i\right)^2 ,
\quad
{\cal L}_{\rm YM}^{(a)}
  =-\frac14\left({\cal F}_{\mu\nu}^a\right)^2 .
\end{equation}

\par
Next, we introduce the auxiliary tensor field $B_{\mu\nu}^i$ of rank two in such a way that the ${\cal L}_{\rm YM}^{(i)}$ piece is recovered after the $B_{\mu\nu}^i$ is integrated out.
Suppose $B_{\mu\nu}^i$ is  anti-symmetric in the indices $\mu$ and $\nu$ and takes values in the Cartan subalgebra; that is, $B_{\mu\nu}^i$ is an anti-symmetric {\it Abelian} tensor field.\footnote{We could start from the BF-YM theory by introducing the {\it non-Abelian} anti-symmetric tensor ${\cal B}_{\mu\nu}$ as the Hodge dual of ${\cal F}_{\mu\nu}$, as discussed in Refs.~\citen{MZ97} and \citen{KondoI}.  However, this approach does not lead to a renormalizable result in the sense explained below.
}
We would like to identify $B_{\mu\nu}^i$ with the Hodge dual%
\footnote{The definition of the Hodge dual (\ref{eq:Hodge}) used in this paper differs from the conventional one by the factor of the imaginary unit $i\equiv\sqrt{-1}$.}
of the Abelian tensor field $Q_{\mu\nu}^i$,
\begin{equation}
 {}^*\!Q_{\mu\nu}^i := {i \over 2} \epsilon_{\mu\nu\rho\sigma} Q^{\rho\sigma}{}^i ,
\label{eq:Hodge}
\end{equation}
 which is assumed to be a composite operator constructed from the gauge potential ${\cal A}_\mu^A$, i.e.,
\begin{equation}
   B_{\mu\nu}^i   =   i {}^*\!Q_{\mu\nu}^i .
\label{eq:dual}
\end{equation}
Therefore, $Q_{\mu\nu}^i$ should be a $U(1)^{N-1}$ invariant anti-symmetric tensor written in terms of the gauge potential ${\cal A}_\mu^A$.
The simplest choice for $Q_{\mu\nu}^i$ is  a linear combination of $f_{\mu\nu}^i$ and $f^{ibc}A_\mu^bA_\nu^c$ with two parameters $\rho$ and $\sigma$, since $Q_{\mu\nu}^i$ is expected to be at most quadratic in the gauge potential from the renormalizability:
\begin{eqnarray}
Q_{\mu\nu}^i
 &=&\rho f_{\mu\nu}^i
    +\sigma gf^{ibc}A_\mu^bA_\nu^c
    \nonumber\\
 &=&\rho(\partial_\mu A_\nu^i
         -\partial_\nu A_\mu^i)
    +\sigma gf^{ibc}A_\mu^bA_\nu^c .
\label{eq:Qdef}
\end{eqnarray}
The advantages of this choice will be clarified shortly.
The diagonal piece of the YM Lagrangian is expanded as
\begin{eqnarray}
{\cal L}_{\rm YM}^{(i)}
 &=&-\frac14\left[f_{\mu\nu}^i
                  +gf^{ibc}A_\mu^bA_\nu^c\right]^2
    \nonumber\\
 &=&-\frac14\left(f_{\mu\nu}^i\right)^2
    -\frac12gf_{\mu\nu}^if^{ibc}A^{\mu b}A^{\nu c}
    -\frac14{g^2}\left(f^{ibc}A_\mu^bA_\nu^c\right)^2 .
\end{eqnarray}
Thus the simplest form satisfying (\ref{eq:dual}) and (\ref{eq:Qdef}) is given by
\begin{eqnarray}
{\cal L}_{\rm YM}^{(i)} &=&-\frac{1-\rho^2}4\left(f_{\mu\nu}^i\right)^2
    -\frac{1-\rho\sigma}2gf_{\mu\nu}^if^{ibc}A^{\mu b}A^{\nu c}
    -\frac{1-\sigma^2}4g^2\left(f^{ibc}A_\mu^bA_\nu^c\right)^2
    \nonumber\\
 & &-\frac14\left(B_{\mu\nu}^i\right)^2
    +\frac i2B_{\mu\nu}^i{}^\ast\!Q^{\mu\nu i} .
\label{eq:L_inv^i}
\end{eqnarray}
In particular, when $\rho=\sigma$, $Q_{\mu\nu}^i$ is nothing but the diagonal component of the non-Abelian field strength, 
$Q_{\mu\nu}^i = \rho {\cal F}_{\mu\nu}^i$, and hence
$B_{\mu\nu}^i   =i\rho {}^*\!{\cal F}_{\mu\nu}^i$.
In view of this, the choice (\ref{eq:Qdef}) is a generalization of that of the previous paper,\cite{KondoI} in which two special cases, 
$\rho=\sigma=1$ and $\rho=0, \sigma=1$, are discussed (Eqs.~(2.9) and (2.12), respectively).\cite{KondoI}
The latter case was first discussed by Quandt and Reinhardt.\cite{QR98}

\par
On the other hand, by defining the covariant derivative with respect to the Abelian gauge field,
\begin{equation}
D_\mu{\mit\Phi}^A
  :=\left(\partial_\mu\delta^{AB}
         +gf^{AiB}A_\mu^i\right){\mit\Phi}^B ,
\end{equation}
the off-diagonal piece can be rewritten as
\begin{equation}
{\cal L}_{\rm YM}^{(a)}
  =-\frac14\left[D_\mu A_\nu^a
                 -D_\nu A_\mu^a
                 +gf^{abc}A_\mu^bA_\nu^c\right]^2 .
\label{eq:L_inv^a}
\end{equation}

\subsection{Maximal Abelian gauge fixing} %

By making use of the Becchi-Rouet-Stora-Tyutin (BRST) transformation $\mbox{\boldmath$\delta$}_{\rm B}$, the gauge fixing and Faddeev-Popov (FP) ghost term is obtained as
\begin{eqnarray}
{\cal L}_{\rm GF+FP} =-i\mbox{\boldmath$\delta$}_{\rm B}G ,
\quad
G &=& \bar C^a\left(F^a+\frac\alpha2\phi^a\right)
  +\bar C^i\left(F^i+\frac\beta2\phi^i\right) ,
\end{eqnarray}
for the gauge fixing condition  $F^a$ of the off-diagonal piece and $F^i$ of the diagonal piece with gauge fixing parameters $\alpha$ and $\beta$, where $\bar C^A$ is the anti-ghost field and $\phi^A$ is the Nakanishi-Lautrup (NL) Lagrange multiplier field.  We adopt the MA gauge for the off-diagonal piece,
\begin{equation}
F^a=D^\mu A_\mu^a ,
\end{equation}
whereas the Lorentz gauge is chosen for the diagonal piece,
\begin{equation}
F^i=\partial^\mu A_\mu^i .
\end{equation}
Here, the BRST transformation is given by
\begin{eqnarray}
\mbox{\boldmath$\delta$}_{\rm B} A_\mu^A
  &=& ({\cal D}_\mu C)^A ,
\nonumber\\
\mbox{\boldmath$\delta$}_{\rm B} C^A &=& -\frac g2f^{ABC}C^BC^C ,
\nonumber\\
\mbox{\boldmath$\delta$}_{\rm B} \bar C^A &=& i\phi^A ,
\nonumber\\
\mbox{\boldmath$\delta$}_{\rm B} \phi^A &=& 0 .
\end{eqnarray}
It should be remarked that the covariant derivative $D_\mu:=D_\mu[A^i]$ is defined for the diagonal gauge field $A_\mu^i$ as
\begin{equation}
D_\mu{\mit\Phi}^a
  :=\left(\partial_\mu\delta^{ab}
         +gf^{aib}A_\mu^i\right){\mit\Phi}^b ,
\end{equation}
 while we define
\begin{equation}
({\cal D}_\mu {\mit\Phi})^A := (\partial_\mu \delta^{AC}+ gf^{ABC} {\cal A}_\mu^B) {\mit\Phi}^C 
= (D_\mu {\mit\Phi})^A + g f^{AbC}A_\mu^b {\mit\Phi}^C . 
\end{equation}
Thus we obtain
\begin{eqnarray}
{\cal L}_{\rm GF+FP}
 &=&\frac\alpha2(\phi^a)^2
    +\phi^aF^a
    +i\bar C^aD^2C^a
    -ig^2f^{abi}f^{icd}\bar C^aA_\mu^bA^{\mu c}C^d    
\nonumber\\
 & &+igf^{abc}\bar C^aD^\mu (A_\mu^bC^c)
    +i\bar C^agf^{abi}F^bC^i
 \nonumber\\
 & &+\frac\beta2(\phi^i)^2
    +\phi^iF^i
    +i\bar C^i\partial^2C^i
    +i\bar C^i\partial^\mu(gf^{ibc}A_\mu^bC^c) .
\label{eq:L_GF+FP}
\end{eqnarray}

\section{The meaning of the renormalizability of the APEGT} %
In order to confirm the necessity of the new parameters
$\rho$ and $\sigma$ introduced in Eq.~(\ref{eq:Qdef}),
we first deal with the case of choosing the parameters
$\rho=0$ and $\sigma=1$ at the tree level of the
Lagrangian~(\ref{eq:L_inv^i}).
Moreover, to simplify the discussion, we consider the $SU(2)$ case.
Then our Lagrangian is given by
\begin{equation}
{\cal L}
 ={\cal L}_{\rm inv}[a_\mu,B_{\mu\nu},A_\mu^a]
  +{\cal L}_{\rm GF+FP}[a_\mu,A_\mu^a,\bar C^A,C^A]
\label{eq:total L},
\end{equation}
where ${\cal L}_{\rm inv}$ is the gauge invariant part of the Lagrangian,
\begin{eqnarray}
{\cal L}_{\rm inv}
 &=&-\frac14\left(f_{\mu\nu}\right)^2
    -\frac14\left(B_{\mu\nu}\right)^2
    -\frac12g\epsilon^{ab}(f-i{}^\ast\!B)_{\mu\nu}A^{\mu a}A^{\nu b}
    \nonumber\\
 & &-\frac14\left(D_\mu A_\nu^a-D_\nu A_\mu^a\right)^2,
\end{eqnarray}
and ${\cal L}_{\rm GF+FP}$ is the gauge fixing and FP ghost term,
\begin{eqnarray}
{\cal L}_{\rm GF+FP}
 &=&\frac\alpha2(\phi^a)^2
    +\phi^a(D^\mu A_\mu^a)
    +\frac\beta2(\phi)^2
    +\phi(\partial^\mu a_\mu)
    \nonumber\\
 & &+i\bar C^aD^2C^a
    -ig^2\epsilon^{ab}\epsilon^{cd}\bar C^aA_\mu^b A^{\mu c}C^d
    \nonumber\\
 & &+i\bar C^ag\epsilon^{ab}(D^\mu A_\mu^b)C^3
    + i\bar C^3 \partial^2 C^3
    +i\bar C^3\partial^\mu(g\epsilon^{ab}A_\mu^aC^b).
\label{eq:SU(2)GFFP}
\end{eqnarray}
Since there is only one diagonal component in the $SU(2)$ gauge
group, we omit the diagonal index and replace the structure constant $f^{ABC}$ by an anti-symmetric tensor $\epsilon^{ab}:=\epsilon^{ab3}$.
\par
It may seem that this theory with the total Lagrangian~(\ref{eq:total L}) is renormalizable, because it appears to be equivalent to the ordinary Yang-Mills theory,
since $B_{\mu\nu}$ is the auxiliary field.
However, this is not the case.
Actually, we will show that undesirable divergent
terms which are absent in the original Lagrangian are induced
as quantum effects; that is,
the renormalizability of this theory is spoiled.
For this reason, we should modify this theory by requiring renormalizability.

First, we consider the gauge invariant part, ${\cal L}_{\rm inv}$.
There are two kinds of undesirable divergence in this theory.
One of them comes from the quartic off-diagonal gluon interaction
through the process (a) of Fig.~\ref{fig:undesirable divergence}.
The other is the bilinear term of the diagonal gluon $a_\mu$
and the auxiliary field $B_{\mu\nu}$ represented by the process (b) of
Fig.~\ref{fig:undesirable divergence}.
\unitlength=0.001in
\begin{figure}[t]
\begin{center}
\begin{picture}(4200,2000)
\put(0,2000){\mbox{(a)}}%
\put(250,900){\epsfysize=30mm \epsfbox{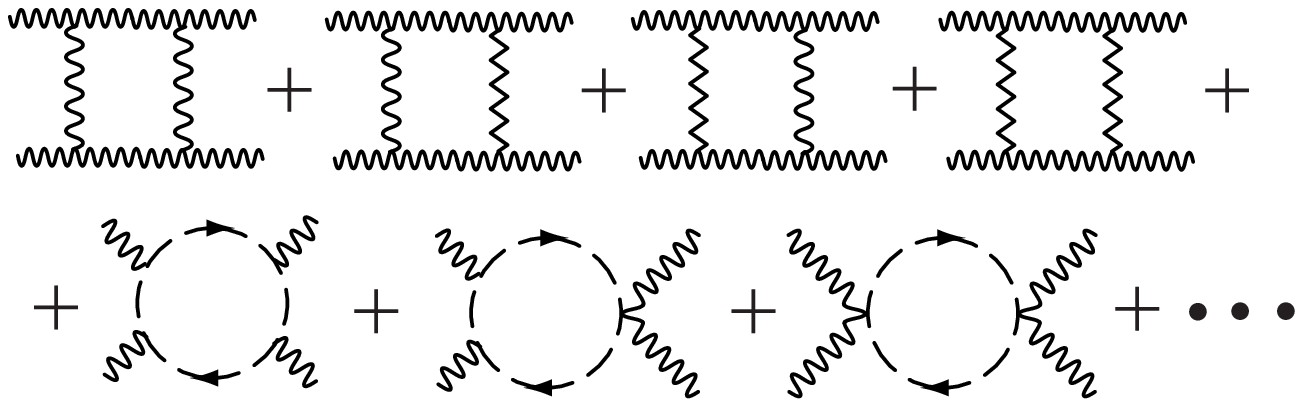}}%
\put(0,550){\mbox{(b)}}%
\put(300,50){\epsfysize=15mm \epsfbox{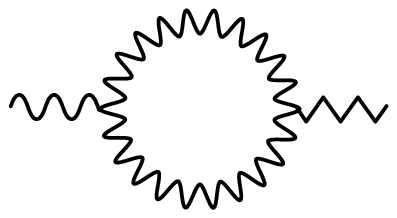}}%
\put(1700,550){\mbox{(c)}}%
\put(2000,70){\epsfysize=15mm \epsfbox{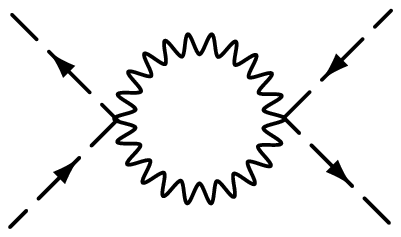}}%
\end{picture}
\end{center}
\caption[]{%
Graphs corresponding to undesirable (nonrenormalizable) divergences. 
Here (a) is the quartic gluon interaction, (b) is the bilinear interaction
of $a_\mu$ and $B_{\mu\nu}$, and (c) is the quartic ghost interaction.
The slowly and rapidly vibrating wavy lines correspond respectively to
the diagonal gluon $a_\mu$ and to the off-diagonal gluon $A_\mu^a$,
while the zig-zag line corresponds to the auxiliary field $B_{\mu\nu}$.
The broken line denotes the ghost or anti-ghost.}
\label{fig:undesirable divergence}
\end{figure}
Neither of these interactions exists in the original
Lagrangian~(\ref{eq:total L}), and hence we cannot absorb such divergences
into the original theory by means of any
renormalization procedure.

Even though we choose the parameters $\rho$ and $\sigma$
appropriately so that the above two interaction terms are included
in the theory from the beginning, we cannot absorb all
divergences into the original theory.
The reason is as follows.
The gauge invariant part of the Lagrangian
(obtained by choosing the parameters $\rho$ and $\sigma$ appropriately),
\begin{eqnarray}
{\cal L}_{\rm inv}^\prime
 &=&-\frac{1-\rho^2}4\left(f_{\mu\nu}\right)^2
    -\frac{1-\rho\sigma}2g\epsilon^{ab}f_{\mu\nu}A^{\mu a}A^{\nu b}
    -\frac{1-\sigma^2}4g^2
     \left(\epsilon^{ab}A^{\mu a}A^{\nu b}\right)^2
    \nonumber\\
 & &-\frac14\left(B_{\mu\nu}\right)^2
    -\frac12g{}^\ast\!B_{\mu\nu}
     \left(\rho f^{\mu\nu}+\sigma \epsilon^{ab}A^{\mu a}A^{\nu b}\right)
    \nonumber\\
 & &-\frac14\left(D_\mu A_\nu^a-D_\nu A_\mu^a\right)^2,
\end{eqnarray}
possesses the interaction terms that we want.
However, as compared with the original $SU(2)$ gauge theory,
three more divergences $\langle B_{\mu\nu}B_{\xi\eta}\rangle$,
$\langle a_\mu B_{\xi\eta}\rangle$
and  $\langle A_\mu^aA_\nu^bB_{\xi\eta}\rangle$  are induced
by introducing the auxiliary field $B_{\mu\nu}$.
Therefore we cannot absorb completely the three new divergences
only in the auxiliary field $B_{\mu\nu}$.
Of course, if there are some relations between these three divergences,
like the Ward identity, it may be possible to do so.
However, regrettably, as shown explicitly in the next section,
there is no such relation, at least between two
propagators $\langle B_{\mu\nu}B_{\xi\eta}\rangle$ and
$\langle a_\mu B_{\xi\eta}\rangle$.
Therefore we must take into account the renormalization of the two
additional parameters $\rho$ and $\sigma$ to preserve the
renormalizability of our theory so that we have totally three
renormalizable quantities together with $B_{\mu\nu}$.
Thus it is possible to absorb all divergences and we can obtain
a renormalizable Lagrangian.
In other words, by taking account of not only the renormalization
of the auxiliary fields $B_{\mu\nu}$ but also that of the two
parameters $\rho$ and $\sigma$, the renormalizability of our
original Lagrangian is preserved.

Next, we proceed to examine the gauge fixing part $S_{\rm GF+FP}$.
Because of the non-linearity of the MA gauge,
this gauge fixing part~(\ref{eq:SU(2)GFFP}) has a significant
distinction from that in the covariant gauge,
the existence of the $\bar CCAA$ interaction, which is the fourth
term of the integrand of Eq.~(\ref{eq:L_GF+FP}).
Since this interaction is shown to induce a divergent quartic ghost
interaction term through the process (c) in
Fig.~\ref{fig:undesirable divergence}, in spite of the absence of
such terms in the original Lagrangian, we should introduce the quartic
ghost interaction term from the beginning.
The way of introducing such an interaction is not unique, and hence we have
proposed the modified MA gauge fixing condition in Ref.~\citen{KS00a}.
According to the modified MA gauge fixing,
the gauge fixing term~(\ref{eq:SU(2)GFFP}) is rewritten as
\begin{eqnarray}
{\cal L}_{\rm GF+FP}^\prime
 &=&\frac\alpha2(\phi^a)^2
    +\phi^a(D^\mu A_\mu^a)
    +\frac\beta2(\phi)^2
    +\phi(\partial^\mu a_\mu)
    \nonumber\\
 & &+i\bar C^aD^2C^a
    -ig^2\epsilon^{ab}\epsilon^{cd}\bar C^aA_\mu^b A^{\mu c}C^d
    \nonumber\\
 & &+i\bar C^ag\epsilon^{ab}(D^\mu A_\mu^b)C^3
    + i\bar C^3 \partial^2 C^3
    +i\bar C^3\partial^\mu(g\epsilon^{ab}A_\mu^aC^b).
    \nonumber\\
 & &-\zeta g\epsilon^{ab}i\phi^a\bar C^bC^3
    +\frac\zeta4g^2\epsilon^{ab}\epsilon^{cd}\bar C^a\bar C^bC^cC^d.
\label{eq:SU(2)GFFP'}
\end{eqnarray}
The only difference between the ordinary MA
gauge~(\ref{eq:SU(2)GFFP}) and the modified
one~(\ref{eq:SU(2)GFFP'}) is the last two
terms in the latter.
Owing to the existence of the quartic ghost interaction term in
the original Lagrangian, the gauge fixing part of the Lagrangian
restores its renormalizability.
\par
Moreover, the introduction of the quartic ghost interaction is
very important from a more physical viewpoint.
When a quartic ghost interaction exists, we find that ghost
condensation occurs.
Then the off-diagonal gluons and the off-diagonal ghosts acquire
their masses in the ghost condensed phase.
Therefore, the Abelian dominance which was previously postulated is
understood as a natural result of these phenomena.
However, we do not give further explanation of
the ghost self-interaction terms here, because they do not
affect the results obtained in this paper.
More detailed discussion is given in Ref.~\citen{KS00a}.
\par
Since our starting Lagrangian
${\cal L}={\cal L}_{\rm inv}^\prime+{\cal L}_{\rm GF+FP}^\prime$
is renormalizable
after taking account of renormalizations of the auxiliary
field $B_{\mu\nu}$ and the two parameters $\rho$ and $\sigma$,
the $\beta$-function and the anomalous dimensions are well-defined,
so that we can obtain them unambiguously.
Because of the existence of quantum corrections of
$B_{\mu\nu}$, $\rho$ and $\sigma$, quantum corrections to
propagators and vertices are different from those in the ordinary
Yang-Mills theory.
However, we expect that the individual anomalous dimensions of the quantities
included in the original Yang-Mills theory, in particular the
$\beta$-function, remain the same.
In fact, after the calculations in the next section, the resulting
$\beta$-function, the anomalous dimensions of diagonal
gluon, $\gamma_a$ and those of the Abelian gauge fixing parameter
$\gamma_\beta$ are seen to be the same as the original ones.  
Thus these results obtained in the next section are far from trivial.
\par
We define the APEGT as an Abelian gauge theory which is expressed in terms of only the diagonal components $a_\mu^i$ and  $B_{\mu\nu}^i$ (and diagonal ghosts $\bar C^i$ and $C^i$).  Therefore, in order to obtain the APEGT, the off-diagonal components $A_\mu^a, \bar C^a$ and $C^a$ must be integrated out in a certain sense, which is discussed in the next section.
Hence, the ``renormalizability'' of the APEGT implies that all 
the divergent terms made of only diagonal fields in the APEGT can be removed consistently after the prescription of the renormalization.
Therefore, in order to obtain a renormalizable APEGT,
we have only to worry about the contribution of the divergent graphs
with only diagonal fields as external legs.
Of course, miscellaneous terms with off-diagonal fields and 
higher derivative terms are generated as quantum effects,
but we are concerned with the situation in which such terms are
not relevant (see Ref.~\citen{Kondo00} for more details).
According to Abelian dominance, we can expect that
such a situation is realized in the low-energy regime of QCD.  
In fact, Abelian dominance is understood as a consequence of dynamical mass generation of off-diagonal gluons and off-diagonal ghosts, as shown in Ref.~\citen{KS00a}.

\section{Renormalization and renormalization-group functions} %
\subsection{The total Lagrangian and the $U(1)^{N-1}$ symmetry} %
The total Lagrangian is obtained by summing up (\ref{eq:L_inv^i}), (\ref{eq:L_inv^a}) and (\ref{eq:L_GF+FP}):
\begin{eqnarray}
{\cal L}
 &=&-\frac{1-\rho^2}4\left(f_{\mu\nu}^i\right)^2
    -\frac{1-\rho\sigma}2gf_{\mu\nu}^if^{ibc}A^{\mu b}A^{\nu c}
    -\frac{1-\sigma^2}4g^2\left(f^{ibc}A_\mu^bA_\nu^c\right)^2
    \nonumber\\
 & &-\frac14\left(B_{\mu\nu}^i\right)^2
    +\frac i2B_{\mu\nu}^i{}^\ast\!Q^{\mu\nu i}
    -\frac14\left[D_\mu A_\nu^a
                 -D_\nu A_\mu^a
                 +gf^{abc}A_\mu^bA_\nu^c\right]^2
    \nonumber\\
 & &+\frac\alpha2(\phi^a)^2
    +\phi^aF^a
    +\frac\beta2(\phi^i)^2
    +\phi^iF^i
    \nonumber\\
 & &+i\bar C^aD^2C^a
    -ig^2f^{abi}f^{icd}\bar C^aA_\mu^bA^{\mu c}C^d
    +igf^{abc}\bar C^aD^\mu(A_\mu^bC^c)
    \nonumber\\
 & &+i\bar C^i\partial^2C^i
    +i\bar C^i\partial^\mu(gf^{ibc}A_\mu^bC^c)
    +i\bar C^agf^{abi}F^bC^i .
\label{eq:L} 
\end{eqnarray}
We separate each field ${\mit\Phi}$ into the background (classical field) $\bar {\mit\Phi}$ and the quantum fluctuation field $\tilde {\mit\Phi}$ as
${\mit\Phi} = \bar {\mit\Phi} + \tilde {\mit\Phi}$.
We then have
$$
{\cal A}_\mu=\bar{\cal A}_\mu+\tilde{\cal A}_\mu,\ 
B_{\mu\nu}=\bar B_{\mu\nu}+\tilde B_{\mu\nu},\ 
\phi=\bar\phi+\tilde\phi,\ 
C=C_{\rm cl}+\tilde C,\ 
\bar C=\bar C_{\rm cl}+\tilde{\bar C} .
$$
(We have used different notation for the ghost and anti-ghost field to avoid unnecessary confusion.)
Here the background field $\bar {\mit\Phi}$ is assumed to satisfy the equation of motion.\footnote{
This standpoint is different from that adopted in the previous paper.\cite{KondoI}
However, it is always possible to translate between the results obtained there and here.
(See the footnotes below.)}
Substituting this decomposition into (\ref{eq:L}), we find that the terms linear in the fluctuation field $\tilde {\mit\Phi}$ vanish due to the equation of 
motion%
\footnote{
Even if $\bar {\mit\Phi}$ does not satisfy the equation of motion, the cross term between the diagonal field and the off-diagonal field can disappear, as
$
 {\rm tr}[\tilde {\mit\Phi}^a T^af(\bar {\mit\Phi}^i) T^i]  =  \tilde {\mit\Phi}^a f(\bar {\mit\Phi}^i)  {\rm tr}(T^a  T^i) = 0.
$
This is the case for the MA gauge, where the background field has a diagonal component alone, and the fluctuation field is given by the off-diagonal components alone, i.e., 
$
 \bar {\mit\Phi} = {\mit\Phi}^i T^i, \tilde {\mit\Phi} = {\mit\Phi}^a T^a .
$
}
\begin{equation}
 {\delta S[\bar {\mit\Phi}] \over \delta \bar {\mit\Phi}} = 0,
\end{equation}
since
\begin{equation}
 S[\bar {\mit\Phi}+\tilde {\mit\Phi}]=S[\bar {\mit\Phi}] + \tilde {\mit\Phi} {\delta S[{\mit\Phi}] \over \delta {\mit\Phi}}\Biggr|_{{\mit\Phi}=\bar {\mit\Phi}} + O(\tilde {\mit\Phi}^2) .
\end{equation}
By taking into account
\begin{equation}
(D_\mu{\mit\Phi})^a
  =\partial_\mu{\mit\Phi}^a
   +gf^{aib}(\bar A+\tilde A)_\mu^i{\mit\Phi}^b
  = (\bar D_\mu{\mit\Phi})^a
   +gf^{aib}\tilde A_\mu^i{\mit\Phi}^b
\end{equation}
and
\begin{equation}
\partial_\mu{\mit\Phi}^i
  = (\bar D_\mu{\mit\Phi})^i ,
\end{equation}
it turns out that the derivative $\partial_\mu$  in the Lagrangian can be replaced by the covariant derivative $\bar D_\mu :=D_\mu[\bar A^i]$ with respect to the background Abelian gauge field $\bar A^i$.
After this replacement, the background Abelian gauge field $\bar A_\mu^i$ appears only in the gauge fixing term $\bar\phi^i\partial_\mu\bar A_\mu^i$ for the background Abelian gauge field $\bar A_\mu^i$.
Therefore, ${\cal L}-\bar\phi^i\partial_\mu\bar A_\mu^i$ is invariant under the following $U(1)^{N-1}$ rotation:
\begin{equation}
\left\{
\begin{array}{l}
\delta\bar A_\mu^i=\partial_\mu\theta^i, \cr
\delta{\mit\Phi}^i=0,
  \quad({\mit\Phi}=\bar B,\bar\phi,C_{\rm cl},\bar C_{\rm cl},
                   \tilde A,\tilde B,\tilde\phi,\tilde C,\tilde{\bar C}) \cr
\delta{\mit\Phi}^a=-gf^{abi}{\mit\Phi}^b\theta^i.
  \quad({\mit\Phi}=\bar A,\bar\phi,C_{\rm cl},\bar C_{\rm cl},
                   \tilde A,\tilde\phi,\tilde C,\tilde{\bar C}) \cr
\end{array}
\right.
\end{equation}
That is to say, under the residual gauge transformation, all the diagonal fields but $\bar A_\mu^i$ are invariant, while an off-diagonal field behaves as a charged matter field.
(Note that $B_{\mu\nu}$ has no off-diagonal component.)
Even after having performed the integration over the fluctuation fields,  
${\cal L}_{\rm eff}-\bar\phi^i\partial_\mu\bar A_\mu^i$ has a residual symmetry given by
\begin{equation}
\left\{
\begin{array}{l}
\delta\bar A_\mu^i=\partial_\mu\theta^i, \cr
\delta{\mit\Phi}^i=0,
  \quad({\mit\Phi}=\bar B,\bar\phi,C_{\rm cl},\bar C_{\rm cl},) \cr
\delta{\mit\Phi}^a=-gf^{abi}{\mit\Phi}^b\theta^i.
  \quad({\mit\Phi}=\bar A,\bar\phi,C_{\rm cl},\bar C_{\rm cl}) \cr
\end{array}
\right.
\end{equation}
The resulting effective Lagrangian reads
\begin{eqnarray}
{\cal L}_{\rm eff}-\bar\phi^i\partial_\mu\bar A_\mu^i
 &=&-\frac{Z_{A^i}^{-1}}4\left(\bar f_{\mu\nu}^i\right)^2
    -\frac{Z_B^{-1}}4\left(\bar B_{\mu\nu}^i\right)^2
    +\frac i2{Z_{AB}^{-1}}\bar B_{\mu\nu}^i
       {}^\ast\!\bar f^{\mu\nu i}
    \nonumber\\
 & &-\frac{Z_{A^a}^{-1}}4\left(\bar D_\mu\bar A_\nu^a
                           -\bar D_\nu\bar A_\mu^a\right)^2
    +\cdots .
\end{eqnarray}
\par
We can observe that the renormalization for $g$ and $\bar{\cal A}_\mu^A$, 
\begin{equation}
g=Z_gg_{\rm R},\quad
\bar A_\mu^i=Z_{A^i}^{1/2}(\bar A_{\rm R})_\mu^i,\quad
\bar A_\mu^a=Z_{A^a}^{1/2}(\bar A_{\rm R})_\mu^a ,
\end{equation}
leads to
\begin{eqnarray}
\bar D_\mu\bar A_\nu^a
 &=&\left(\partial_\mu\delta^{ab}
          +gf^{aib}\bar A_\mu^i\right)\bar A_\nu^b
    \nonumber\\
 &=&Z_{A^a}^{1/2}
    \left[\partial_\mu\delta^{ab}
          +Z_gZ_{A^i}^{1/2}g_{\rm R}f^{aib}(\bar A_{\rm R})_\mu^i\right]
    (\bar A_{\rm R})_\nu^b .
\end{eqnarray}
Hence, renormalizability requires the relation
\begin{equation}
  Z_gZ_{A^i}^{1/2}=1.
\label{eq:Z_g-Z_A}
\end{equation}
This relation can also be derived from the Ward-Takahashi identity for the residual $U(1)^{N-1}$ symmetry, which is preserved due to the nature of the background field method,\cite{Abbott82} as has been demonstrated in the $SU(2)$ case ($N=2$).\cite{KondoI}

\subsection{General consideration} %

If we perform the functional integration over the fluctuation fields, a divergence appears as a quantum effect.  This divergence can be removed by renormalization of the fields ($\bar{\mit\Phi},\tilde{\mit\Phi}$), the coupling constant $g$, and the two parameters $\rho$ and $\sigma$.
To simplify the argument, we first consider the divergence appearing in the sub-graph with external lines $\tilde{\mit\Phi}$.
Since $\bar{\mit\Phi}$ does not appear in the internal lines, we can set $\bar{\mit\Phi}=0$.  Hence we consider $\tilde{\cal L}$, which is obtained from  (\ref{eq:L}), by replacing ${\mit\Phi}$ with $\tilde {\mit\Phi}$.
After integrating over  $\tilde B$ and $\tilde\phi$ in $\tilde{\cal L}$, we obtain
\begin{eqnarray}
\tilde{\cal L}
 &=&-\frac14\left(\tilde {\cal F}_{\mu\nu}^A\right)^2
+ \tilde{\cal L}_{GF+FP} ,
\\
\tilde{\cal L}_{GF+FP} &:=&
    -\frac1{2\alpha}\left(\tilde F^a\right)^2
    -\frac1{2\beta}\left(\tilde F^i\right)^2
    \nonumber\\
 & &+i\tilde{\bar C}^a\tilde D^2\tilde C^a
    -ig^2f^{abi}f^{icd}\tilde{\bar C}^a\tilde A_\mu^b
                       \tilde A^{\mu c}\tilde C^d
    +igf^{abc}\tilde{\bar C}^a\tilde D^\mu(\tilde A_\mu^b\tilde C^c)
    \nonumber\\
 & &+i\tilde{\bar C}^i\partial^2\tilde C^i
    +i\tilde{\bar C}^i\partial^\mu(gf^{ibc}\tilde A_\mu^b\tilde C^c)
    +i\tilde{\bar C}^agf^{abi}\tilde F^b\tilde C^i ,
\end{eqnarray}
where $\tilde D_\mu:=D_\mu[\tilde A_\mu^i]$.
Thus we find that $\tilde{\cal L}$ depends on neither $\rho$ nor $\sigma$.
It can be shown that the divergence can be absorbed into the renormalization of $\tilde{\cal A}_\mu^A$, $\tilde C^A$, $\tilde{\bar C}{}^A$ and $g$ by multiplicative renormalization.  
Then the renormalization constant $Z_g$ of $g$ is independent of $\rho$ and $\sigma$, and hence the $\beta$-function is also independent of $\rho$ and $\sigma$.
The simplest way to determine the renormalization of $\tilde{\cal A}_\mu^A$, $\tilde C^A$, $\tilde{\bar C}{}^A$ and $g$ is to consider the renormalization of the propagators of $\tilde{\cal A}_\mu^A$, $\tilde C^A$ and $\tilde{\bar C}{}^A$ and of the $\tilde{\cal A}_\mu\tilde C\tilde{\bar C}$-vertex.
Nevertheless, it would be rather difficult to calculate the higher-point vertex when we consider the higher loop effect.  
In any case, $\tilde{\cal L}$ does not depend on $\rho$ and $\sigma$ and therefore we do not have to consider the renormalization of the internal lines hereafter.

Next, we evaluate the graph which has only $\bar A_\mu^i$ and $\bar B_{\mu\nu}^i$ as its external lines.
We do not have to consider the renormalization of internal lines $\tilde{\mit\Phi}$, since $\tilde{\mit\Phi}$ appears only in the internal lines, and the renormalization of $\tilde {\mit\Phi}$ has been performed in the previous step.
This fact is well known in the context of the background field method (see Ref.~\citen{Abbott82}).
We have only to consider the renormalization of $\bar A_\mu^i$, $\bar B_{\mu\nu}^i$, $g$, $\rho$ and $\sigma$.
Setting $\bar {\mit\Phi}^a=0$ (the absence of off-diagonal background fields) 
and integrating over $\tilde B_{\mu\nu}^i$ and $\tilde \phi$
 in the Lagrangian,\footnote{%
This could correspond to the Abelian dominance in the sense that the off-diagonal components are negligible in the low-energy region of QCD. 
In other words, the off-diagonal component does not have a low-energy mode from the viewpoint of the Wilsonian renormalization group.  In contrast to the off-diagonal component, the diagonal component has both high-energy and low-energy modes.  In the present treatment, the high-energy modes of the diagonal and the off-diagonal components are integrated out, although the integration of the diagonal high-energy mode was ignored in the previous treatment.\cite{KondoI} However, the result is unchanged, at least at the one-loop level (see Ref.~\citen{KS00c} for more details).}
we obtain
\begin{eqnarray}
{\cal L}
 &=& \bar{\cal L}_c
    -\frac{1-\rho\sigma}2g
     \bar f_{\mu\nu}^if^{ibc}\tilde A^{\mu b}\tilde A^{\nu c}
    -\frac14\sigma g\epsilon^{\mu\nu\rho\sigma}
     \bar B_{\mu\nu}^if^{ibc}\tilde A_\rho^b\tilde A_\sigma^c
    \nonumber\\
 & &-\frac14\left(\bar D_\mu\tilde A_\nu^a
                  -\bar D_\nu\tilde A_\mu^a\right)^2
    -gf^{abc}\left(\bar D_\mu\tilde A_\nu^a\right)
             \tilde A^{\mu b}\tilde A^{\nu c}
    \nonumber\\
 & &-\frac14\left(\tilde F_{\mu\nu}^A\right)^2
    + \tilde{\cal L}_{\rm GF+FP}[\tilde D \rightarrow D] ,
\end{eqnarray}
where the last term $\tilde{\cal L}_{\rm GF+FP}[\tilde D \rightarrow D]$ is obtained by replacing $\tilde D$ with $D$ from $\tilde{\cal L}_{\rm GF+FP}$ and 
\begin{equation}
\bar{\cal L}_c
  =-\frac{1-\rho^2}4\left(\bar f_{\mu\nu}^i\right)^2
   -\frac14\left(\bar B_{\mu\nu}^i\right)^2
   +\frac i2\rho\bar B_{\mu\nu}^i{}^\ast\!\bar f^{\mu\nu i}
   -\frac1{2\beta}\left(\partial^\mu\bar A_\mu^i\right)^2 .
\label{eq:L0}
\end{equation}

It is not difficult to see that the renormalization of $\bar A_\mu^i$, $\bar B_{\mu\nu}^i$, $g$ and $\rho$ is sufficient at the one-loop level.
In fact, the classical (background field) part is given by $\bar{\cal L}_c$,
which does not contain $\sigma$.  The renormalization of $\sigma$ does appear at the two-loop level.  An advantage of the background field method is that we have only to calculate the vertex function at the one-loop level to know the renormalization of $\sigma$ at the two-loop level.
Furthermore, it is enough to consider the three propagators $\bar A_\mu^i$-$\bar A_\nu^j$, $\bar B_{\mu\nu}^i$-$\bar B_{\xi\eta}^j$ and $\bar A_\mu^i$-$\bar B_{\xi\eta}^j$ in order to obtain the $\beta$-function (at the one-loop level). 
The reason is as follows.
At first glance, we need five relationships that completely specify the renormalization of $\bar A_\mu^i, \bar B_{\mu\nu}^i, g, \rho$ and $\sigma$.  However, we have an additional relation (\ref{eq:Z_g-Z_A}) between the renormalization constants for $\bar A_\mu^i$ and $g$, and we do not have to consider the renormalization of $\sigma$ at the one-loop, as mentioned above.
In order to determine all the renormalization factors, therefore, we need only three independent relations (at least at the one-loop level), which are provided by the renormalization conditions (counterterms) of three propagators, as calculated below.

\subsection{Feynman rules} %
\begin{figure}[b]
\begin{center}
\begin{picture}(4300,2400)
\put(0,2200){\mbox{\large(a)}}%
\put(550,2100){%
   \put(0,0){\epsfysize=5mm \epsfbox{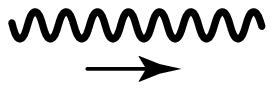}}%
   \put(-280,110){\mbox{$a,\mu$}}%
   \put(700,110){\mbox{$b,\nu$}}%
   \put(230,-50){\mbox{$p$}}%
   }%
\put(1750,2200){\mbox{\large(b)}}%
\put(2200,2200){%
   \put(0,0){\epsfysize=2mm \epsfbox{ghost.eps}}%
   \put(-130,0){\mbox{$a$}}%
   \put(680,0){\mbox{$b$}}%
   \put(250,-100){\mbox{$p$}}%
   }%
\put(0,1700){\mbox{\large(c)}}%
\put(200,1000){%
   \put(0,0){\epsfysize=20mm \epsfbox{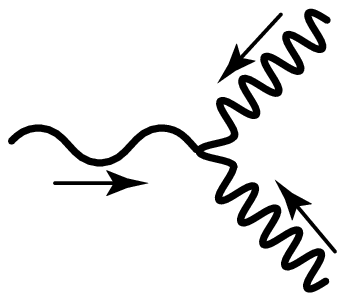}}%
   \put(170,200){\mbox{$p$}}%
   \put(630,750){\mbox{$q$}}%
   \put(900,150){\mbox{$r$}}%
   \put(100,500){\mbox{$i,\mu$}}%
   \put(900,600){\mbox{$a,\rho$}}%
   \put(500,30){\mbox{$b,\sigma$}}%
   }%
\put(1600,1700){\mbox{\large(d)}}%
\put(1800,1000){%
   \put(0,0){\epsfysize=20mm \epsfbox{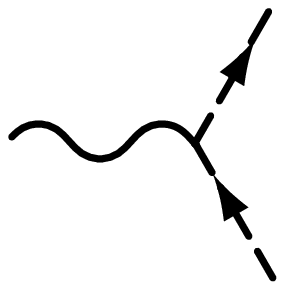}}%
   \put(600,700){\mbox{$p$}}%
   \put(700,150){\mbox{$q$}}%
   \put(100,500){\mbox{$i,\mu$}}%
   \put(750,600){\mbox{$a$}}%
   \put(600,30){\mbox{$b$}}%
   }%
\put(2900,1700){\mbox{\large(e)}}%
\put(3100,1000){%
   \put(0,0){\epsfysize=20mm \epsfbox{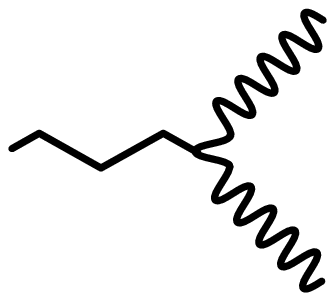}}%
   \put(100,500){\mbox{$i,\mu\nu$}}%
   \put(900,600){\mbox{$a,\rho$}}%
   \put(500,30){\mbox{$b,\sigma$}}%
   }%
\put(0,700){\mbox{\large(f)}}%
\put(400,0){%
   \put(0,0){\epsfysize=20mm \epsfbox{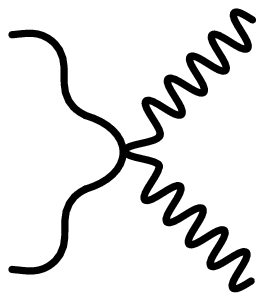}}%
   \put(-150,600){\mbox{$i,\mu$}}%
   \put(-150,100){\mbox{$j,\nu$}}%
   \put(700,600){\mbox{$a,\rho$}}%
   \put(700,100){\mbox{$b,\sigma$}}%
   }%
\put(1600,700){\mbox{\large(g)}}%
\put(2050,0){%
   \put(0,0){\epsfysize=20mm \epsfbox{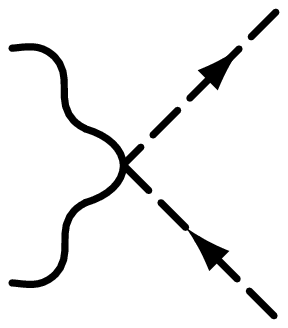}}%
   \put(-150,580){\mbox{$i,\mu$}}%
   \put(-150,130){\mbox{$j,\nu$}}%
   \put(650,600){\mbox{$a$}}%
   \put(650,100){\mbox{$b$}}%
   }%
\end{picture}
\caption[]{%
The graphs in (a) and (b) represent fluctuation-field propagators.
The (rapidly vibrating) wavy line denotes the fluctuation
off-diagonal gluon $\tilde A_\mu^a$, and the broken line denotes
the fluctuation ghost $\tilde C^a$ or anti-ghost $\tilde{\bar C}^a$.
The graphs in (c), (d) and (e) are three-point vertices,
and those in (f) and (g) are four-point vertices.
The (slowly vibrating) wavy line corresponds to the background
diagonal gluon $\bar A_\mu^i$, while the zig-zag line corresponds to the
background anti-symmetric tensor field ${}^\ast\!\bar B_{\mu\nu}^i$.
}
\label{fig:Feynman Rules}
\end{center}
\end{figure}

The Feynman rules are given as follows (see Fig.~\ref{fig:Feynman Rules}).
We give only those rules that are necessary for the renormalization at the one-loop level.
The two-loop result will be given in a subsequent paper.\cite{KS00c}

\subsubsection{Propagators} %
\begin{enumerate}
\item[(a)] Fluctuation off-diagonal gluon propagators: 
\begin{equation}
iD_{\mu\nu}^{ab}
  =-\frac i{p^2}
    \left[g_{\mu\nu}-(1-\alpha)\frac{p_\mu p_\nu}{p^2}\right]\delta^{ab} .
\end{equation}

\item[(b)] Fluctuation off-diagonal ghost propagators: 
\begin{equation}
i\Delta^{ab}
  =-\frac1{p^2}\delta^{ab} .
\end{equation}
\end{enumerate}

\subsubsection{Three-point vertices} %
\begin{enumerate}
\item[(c)] One diagonal and two off-diagonal gluons:
\begin{eqnarray}
&&i\left<\bar A_\mu^i(p)\tilde A_\rho^a(q)
                 \tilde A_\sigma^b(r)\right>_{\rm bare}
  \nonumber\\
&&\textstyle
 =gf^{iab}\left[\!\!\left[ (q-r)_\mu
                +\left\{r-(1-\rho\sigma)p+\frac q\alpha\right\}_\rho
                +\left\{(1-\rho\sigma)p-q-\frac r\alpha\right\}_\sigma
                \right]\!\!\right] ,\quad
\end{eqnarray}
where we have introduced the abbreviated notation
\begin{equation}
[\![A_\mu+B_\rho+C_\sigma]\!]
  =A_\mu g_{\rho\sigma}
   +B_\rho g_{\sigma\mu}
   +C_\sigma g_{\mu\rho} .
\end{equation}

\item[(d)] One diagonal gluon, one off-diagonal ghost and one anti-ghost:
\begin{equation}
i\left<\bar A_\mu^i\tilde{\bar C}^a(p)\tilde C^b(q)\right>_{\rm bare}
  =i(p+q)_\mu gf^{aib} .
\end{equation}

\item[(e)] One diagonal tensor and two off-diagonal gluons:
\begin{equation}
i\left<{}^\ast\!\bar B_{\mu\nu}^i\tilde A_\rho^a
                 \tilde A_\sigma^b\right>_{\rm bare}
  =-\sigma g(g_{\mu\rho}g_{\nu\sigma}-g_{\mu\sigma}g_{\nu\rho})f^{iab} .
\end{equation}

\end{enumerate}

\subsubsection{Four-point vertices} %
\begin{enumerate}
\item[(f)] Two diagonal gluons and two off-diagonal gluons:
\begin{equation}
i\left<\bar A_\mu^i\bar A_\nu^j
       \tilde A_\rho^a\tilde A_\sigma^b\right>_{\rm bare}
  =ig^2f^{aic}f^{cjb}
    \left[2g_{\mu\nu}g_{\rho\sigma}
          -\left(1-\frac1\alpha\right)
          (g_{\mu\rho}g_{\nu\sigma}+g_{\mu\sigma}g_{\nu\rho})\right] .
\end{equation}

\item[(g)] Two diagonal gluons, one off-diagonal ghost and one anti-ghost:
\begin{equation}
i\left<\bar A_\mu^i\bar A_\nu^j
       \tilde{\bar C}^a\tilde C^b\right>_{\rm bare}
  =-2g^2f^{aic}f^{cjb}g_{\mu\nu} .
\end{equation}

\end{enumerate}

\subsection{Counterterms} %
By substituting the following renormalization relations into the Lagrangian (\ref{eq:L0}), 
\begin{eqnarray}
\bar A_\mu^i
 = Z_A^{1/2}(\bar A_{\rm R})_\mu^i,
\quad
\bar B_{\mu\nu}^i
 = Z_B^{1/2}(\bar B_{\rm R})_{\mu\nu}^i,
\quad
\rho= Z_\rho\rho_{\rm R},
\quad
g = Z_g g_R ,
\end{eqnarray}
we obtain
\begin{eqnarray}
\bar{\cal L}_{AB}
 &=&-\frac{1-\rho^2}4\left(\bar f_{\mu\nu}^i\right)^2
    +\frac i2\rho\bar B_{\mu\nu}^i{}^\ast\!\bar f^{\mu\nu i}
    -\frac14\left(\bar B_{\mu\nu}^i\right)^2
    \nonumber\\
 &=&-\frac{1-Z_\rho^2\rho_{\rm R}^2}4
     Z_A\left[(\bar f_{\rm R})_{\mu\nu}^i\right]^2
    +\frac i2Z_\rho Z_A^{1/2}Z_B^{1/2}
     \rho_{\rm R}(\bar B_{\rm R})_{\mu\nu}^i
     {}^\ast\!(\bar f_{\rm R})^{\mu\nu i}
    \nonumber\\
 & &-\frac14Z_B\left[(\bar B_{\rm R})_{\mu\nu}^i\right]^2.
\label{eq:L1}
\end{eqnarray}
On the other hand, the renormalized Lagrangian with the counterterm is written as
\begin{eqnarray}
\bar{\cal L}_{AB}
 &=&-\frac{1-\rho_{\rm R}^2}4\left[(\bar f_{\rm R})_{\mu\nu}^i\right]^2
    +\frac i2\rho_{\rm R}(\bar B_{\rm R})_{\mu\nu}^i
     {}^\ast\!(\bar f_{\rm R})^{\mu\nu i}
    -\frac14\left[(\bar B_{\rm R})_{\mu\nu}^i\right]^2
    \nonumber\\
 & &-\frac{\delta_1}4\left[(\bar f_{\rm R})_{\mu\nu}^i\right]^2
    +\frac i2\delta_2(\bar B_{\rm R})_{\mu\nu}^i
     {}^\ast\!(\bar f_{\rm R})^{\mu\nu i}
    -\frac{\delta_3}4\left[(\bar B_{\rm R})_{\mu\nu}^i\right]^2.
\label{eq:L2}
\end{eqnarray}
By equating (\ref{eq:L1}) and ({\ref{eq:L2}), we find
\begin{eqnarray}
\delta_1
  &=& Z_A^{(1)}-\left(Z_A^{(1)}+2Z_\rho^{(1)}\right)\rho_{\rm R}^2 ,
\nonumber\\
\delta_2
  &=& \left(Z_\rho^{(1)}+\frac12Z_A^{(1)}
         +\frac12Z_B^{(1)}\right)\rho_{\rm R} ,
\nonumber\\
\delta_3 &=& Z_B^{(1)} ,
\label{eq:delta}
\end{eqnarray}
where we have used the expansion of the renormalization factor
$
 Z_{\mit\Phi} = 1+Z_{\mit\Phi}^{(1)} +Z_{\mit\Phi}^{(2)} + \dots ,
$
with $Z_{\mit\Phi}^{(n)}$ being the $n$-loop contribution.
}

\subsection{$\beta$-function and anomalous dimension} %

\begin{figure}[b]
\begin{center}
\begin{picture}(4300,2000)
\put(0,1900){\mbox{(a1)}}%
\put(100,1300){%
   \put(0,0){\epsfysize=17mm\epsfbox{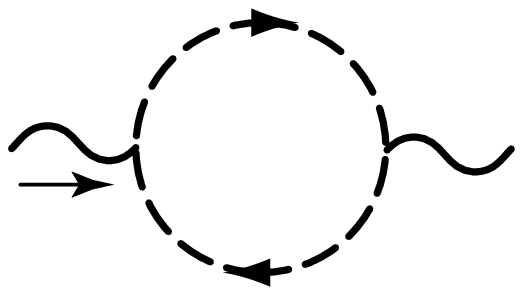}}%
   \put(-80,430){\mbox{$i,\mu$}}%
   \put(1050,400){\mbox{$j,\nu$}}%
   \put(70,130){\mbox{$q$}}%
   \put(550,-80){\mbox{$p$}}%
   \put(420,490){\mbox{$p+q$}}%
   }%
\put(1600,1900){\mbox{(a2)}}%
\put(1700,1250){%
   \put(0,0){\epsfysize=18mm\epsfbox{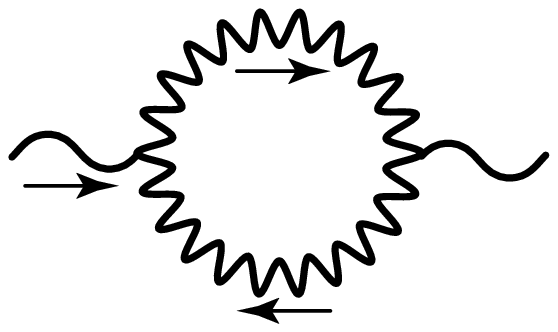}}%
   \put(-80,470){\mbox{$i,\mu$}}%
   \put(1050,420){\mbox{$j,\nu$}}%
   \put(70,220){\mbox{$q$}}%
   \put(620,-50){\mbox{$p$}}%
   \put(420,450){\mbox{$p+q$}}%
   }%
\put(3250,1900){\mbox{(a3)}}%
\put(3470,1250){%
   \put(0,0){\epsfysize=19mm\epsfbox{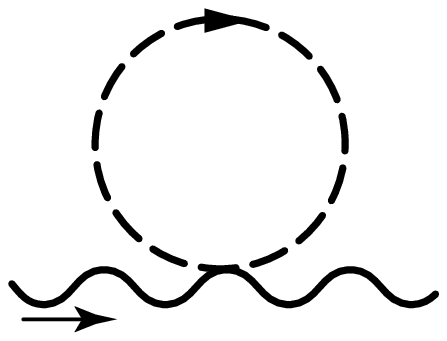}}%
   \put(-100,170){\mbox{$i,\mu$}}%
   \put(900,160){\mbox{$j,\nu$}}%
   \put(450,600){\mbox{$p$}}%
   \put(70,-50){\mbox{$q$}}%
   }%
\put(0,900){\mbox{(a4)}}%
\put(220,130){%
   \put(0,0){\epsfysize=20mm\epsfbox{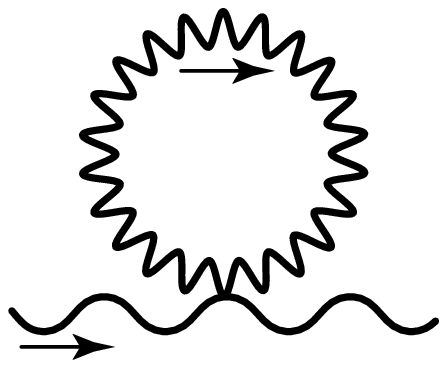}}%
   \put(-100,150){\mbox{$i,\mu$}}%
   \put(900,100){\mbox{$j,\nu$}}%
   \put(420,550){\mbox{$p$}}%
   \put(50,-70){\mbox{$q$}}%
   }%
\put(1600,900){\mbox{(b)}}%
\put(1700,200){%
   \put(0,0){\epsfysize=18mm\epsfbox{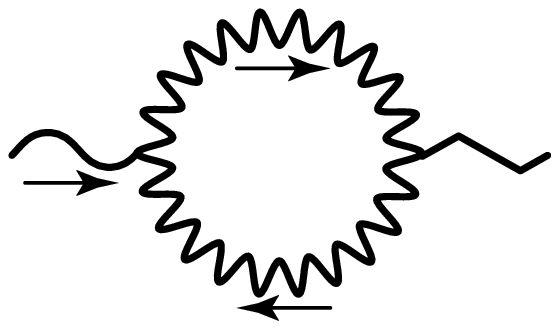}}%
   \put(-80,470){\mbox{$i,\mu$}}%
   \put(1050,450){\mbox{$j,\xi\eta$}}%
   \put(70,220){\mbox{$q$}}%
   \put(620,-50){\mbox{$p$}}%
   \put(420,450){\mbox{$p+q$}}%
   }%
\put(3250,900){\mbox{(c)}}%
\put(3350,200){%
   \put(0,0){\epsfysize=18mm\epsfbox{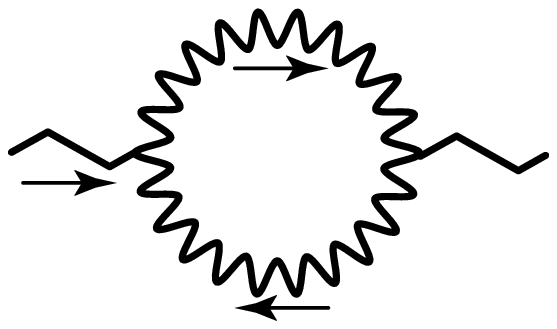}}%
   \put(-80,470){\mbox{$i,\mu\nu$}}%
   \put(1050,450){\mbox{$j,\xi\eta$}}%
   \put(70,220){\mbox{$q$}}%
   \put(620,-50){\mbox{$p$}}%
   \put(420,450){\mbox{$p+q$}}%
   }%
\end{picture}
\end{center}
\caption[]{Vacuum polarization graphs that are necessary to obtain three
           propagators (\ref{eq:counter}) at the one-loop level.}
\label{fig:vacuum polarization}
\end{figure}

The three propagators 
$
\left<(\bar A_{\rm R})_\mu^i(\bar A_{\rm R})_\nu^j\right>
$,
$
\left<(\bar A_{\rm R})_\mu^i
       {}^\ast\!(\bar B_{\rm R})_{\xi\eta}^j\right>
$ and
$ 
\left<{}^\ast\!(\bar B_{\rm R})_{\mu\nu}^i
       {}^\ast\!(\bar B_{\rm R})_{\xi\eta}^j\right>
$
are obtained by calculating the Feynman graphs shown in Fig.~\ref{fig:vacuum polarization}, based on the Feynman rule given in Fig.~\ref{fig:Feynman Rules}.  Consequently, (a1)+(a2)+(a3)+(a4), (b) and (c) give  the counterterms of the respective propagator defined by
\begin{eqnarray}
i\left<(\bar A_{\rm R})_\mu^i(\bar A_{\rm R})_\nu^j\right>_{\rm counter}
  &=& -i\delta_1(q^2g_{\mu\nu}-q_\mu q_\nu)\delta^{ij},
\nonumber\\
i\left<(\bar A_{\rm R})_\mu^i
       {}^\ast\!(\bar B_{\rm R})_{\xi\eta}^j\right>_{\rm counter}
  &=& i\delta_2(q_\xi g_{\mu\eta}-q_\eta g_{\mu\xi})\delta^{ij},
\nonumber\\
i\left<{}^\ast\!(\bar B_{\rm R})_{\mu\nu}^i
       {}^\ast\!(\bar B_{\rm R})_{\xi\eta}^j\right>_{\rm counter}
  &=& -i\delta_3(g_{\mu\xi}g_{\nu\eta}-g_{\mu\eta}g_{\nu\xi})\delta^{ij}
.
\label{eq:counter}
\end{eqnarray}
Straightforward but somewhat tedious calculations employing dimensional regularization determine $\delta_1$, $\delta_2$ and $\delta_3$ as
\begin{eqnarray}
\delta_1
  &=&\delta_{{\rm a1}}+\delta_{{\rm a2}}
=
   \left[(2-\rho_{\rm R}\sigma_{\rm R})^2-\frac13
         +\frac{1-\alpha_{\rm R}}2
          (2-\rho_{\rm R}\sigma_{\rm R})\rho_{\rm R}\sigma_{\rm R}\right]
   \frac{(\mu^{-\epsilon}g_{\rm R})^2}{(4\pi)^2}\frac{C_2(G)}{\epsilon} ,
\nonumber\\
 &&
 \Bigg\{
 \begin{array}{l}
  \displaystyle
  \delta_{{\rm a1}}=
    \frac13
    \frac{(\mu^{-\epsilon}g_{\rm R})^2}{(4\pi)^2}\frac{C_2(G)}{\epsilon} ,
   \cr
  \displaystyle
  \delta_{{\rm a2}}=
   \left[(2-\rho_{\rm R}\sigma_{\rm R})^2-\frac23
         +\frac{1-\alpha_{\rm R}}2
          (2-\rho_{\rm R}\sigma_{\rm R})\rho_{\rm R}\sigma_{\rm R}\right]
   \frac{(\mu^{-\epsilon}g_{\rm R})^2}{(4\pi)^2}\frac{C_2(G)}{\epsilon} ,
 \end{array}
\nonumber\\
\delta_2
  &=&\left[\sigma_{\rm R}(2-\rho_{\rm R}\sigma_{\rm R})
           -\frac{1-\alpha_{\rm R}}2\sigma_{\rm R}
            (1-\rho_{\rm R}\sigma_{\rm R})\right]
     \frac{(\mu^{-\epsilon}g_{\rm R})^2}{(4\pi)^2}
     \frac{C_2(G)}{\epsilon},
\nonumber\\
\delta_3
  &=&-\frac{1+\alpha_{\rm R}}2\sigma_{\rm R}^2
      \frac{(\mu^{-\epsilon}g_{\rm R})^2}{(4\pi)^2}
      \frac{C_2(G)}{\epsilon}
\label{eq:delta2} ,
\end{eqnarray}
where $\delta_{{\rm a1}}$ and $\delta_{{\rm a2}}$ are the contributions from the graphs (a1) and (a2) in Fig.~\ref{fig:vacuum polarization} respectively.%
\footnote{In fact, the contributions from (a1) and (a2) are transverse, as suggested by the background field method.\cite{Abbott82}
By contrast, both contributions from (a3) and (a4) vanish in the dimensional regularization.}
$C_2(G)$ is a quadratic Casimir operator defined by 
$C_2(G) \delta^{AB} = f^{ACD}f^{BCD}$ and $C_2(G)=N$ for $G=SU(N)$.
For a special choice of the parameters, $\rho=\sigma=0$ and $\alpha=0$, $\delta_1$ has been calculated by Quandt and Reinhardt,\cite{QR98} at least for $SU(2)$.
Finally, by equating (\ref{eq:delta}) and (\ref{eq:delta2}), we obtain
\begin{eqnarray}
Z_A^{(1)}
  &=& \frac{11}3\frac{(\mu^{-\epsilon}g_{\rm R})^2}{(4\pi)^2}\frac{C_2(G)}{\epsilon} ,
\nonumber\\
Z_B^{(1)}
  &=& -\frac{1+\alpha_{\rm R}}2\sigma_{\rm R}^2
       \frac{(\mu^{-\epsilon}g_{\rm R})^2}{(4\pi)^2}
       \frac{C_2(G)}{\epsilon} ,
\nonumber\\
Z_\rho^{(1)}
  &=& \left[-\frac{11}6-\frac{\sigma_{\rm R}^2}2
            +2\frac{\sigma_{\rm R}}{\rho_{\rm R}}
            -\frac{1-\alpha_{\rm R}}2
             \left(\frac{\sigma_{\rm R}}{\rho_{\rm R}}
                   -\frac{\sigma_{\rm R}^2}2\right)\right]
   \frac{(\mu^{-\epsilon}g_{\rm R})^2}{(4\pi)^2}\frac{C_2(G)}{\epsilon} .
\end{eqnarray}
The renormalization factor of the coupling constant is obtained from (\ref{eq:Z_g-Z_A}) as
\begin{equation}
Z_g^{(1)}
  =-\frac12Z_A^{(1)}
  =-\frac{11}6\frac{(\mu^{-\epsilon}g_{\rm R})^2}{(4\pi)^2}\frac{C_2(G)}{\epsilon} .
\end{equation}
This implies that the renormalization of $g$ is independent of $\rho, \sigma$ and the gauge parameter $\alpha$.  Therefore, the $\beta$-function is also independent of the choice of $\rho, \sigma$ and the gauge parameter $\alpha$.  The resulting $\beta$-function coincides exactly with the one-loop $\beta$-function of the original $SU(N)$ Yang-Mills theory, 
\begin{equation}
 \beta(g_{\rm R}) := \mu {\partial g_{\rm R} \over \partial \mu} 
 = - g_{\rm R} \mu {\partial \over \partial \mu} \ln Z_g = - {b_0 \over (4\pi)^2} g_{\rm R}^3 + O(g_{\rm R}^5) ,
\quad b_0 = {11 \over 3} C_2(G) ,
\end{equation}
exhibiting asymptotic freedom.
Moreover, the anomalous dimensions of the fields $\bar A_\mu^i$ and $\bar B_{\mu\nu}^i$ and the parameters $\rho$ and $\beta$ are obtained as
\begin{eqnarray}
 \gamma_A(g) &=& {1 \over 2} \mu {\partial \over \partial\mu} \ln Z_A 
=-\frac{11}3 C_2(G) \frac{g_{\rm R}^2}{(4\pi)^2}  ,
\nonumber\\
 \gamma_B(g) &=& {1 \over 2} \mu {\partial \over \partial\mu} \ln Z_B 
=  \frac{1+\alpha_{\rm R}}2\sigma_{\rm R}^2
    C_2(G) \frac{g_{\rm R}^2}{(4\pi)^2} ,
\nonumber\\
\gamma_\rho(g)
 &=& -\rho_{\rm R}\mu\frac\partial{\partial\mu}\ln Z_\rho
\nonumber\\
 &=&-2\rho_{\rm R}
    \left[\frac{11}6+\frac{\sigma_{\rm R}^2}2
          -2\frac{\sigma_{\rm R}}{\rho_{\rm R}}
          +\frac{1-\alpha_{\rm R}}2
           \left(\frac{\sigma_{\rm R}}{\rho_{\rm R}}
                 -\frac{\sigma_{\rm R}^2}2\right)\right]
    C_2(G)\frac{g_{\rm R}^2}{(4\pi)^2} ,
\nonumber\\
 \gamma_\beta(g) &=& -\beta_{\rm R} \mu {\partial \over \partial\mu} \ln Z_A 
= -2 \gamma_A(g) \beta_{\rm R} .
\end{eqnarray}
It turns out that the anomalous dimension depends in general on the gauge parameter $\alpha$.
But $\gamma_A$ is independent of $\alpha$, as expected from the background field method.\cite{Abbott82}

\section{Conclusion and discussion} %
By requiring renormalizability, we have derived a renormalizable
APEGT as a low-energy effective theory of QCD.  The essential part is given by 
\begin{eqnarray}
 {\cal L}_{\rm APEGT}
 = -\frac{1-\rho_{\rm R}^2}4\left[(\bar f_{\rm R})_{\mu\nu}^i\right]^2
    -\frac14\left[(\bar B_{\rm R})_{\mu\nu}^i\right]^2
    +\frac i2\rho_{\rm R}(\bar B_{\rm R})_{\mu\nu}^i
     {}^\ast\!(\bar f_{\rm R})^{\mu\nu i} .
\label{eq:APEGT}
\end{eqnarray}
The coupling constant $g_{\rm R}$ in the APEGT runs according to the $\beta$-function, which is exactly the same as in the original Yang-Mills theory.  The obtained $\beta$-function is independent of the parameters $\rho$ and $\sigma$ and the gauge fixing parameter $\alpha$ of the MA gauge.
\par
The advantages of the renormalizable APEGT are as follows.
Thanks to the renormalizability, the relation between the dual field $B_{\mu\nu}$ and the original field $f_{\mu\nu}$ is preserved after the renormalization:
\begin{equation}
 (\bar B_{\rm R})_{\mu\nu}^i = i
     \rho_{\rm R} {}^\ast\!(\bar f_{\rm R})_{\mu\nu}^i .
\end{equation}
Note that $\sigma$ does not appear in this relation, simply because we have set $\bar A_\mu^a=0$ in the derivation of (\ref{eq:L0}).
Moreover, we can switch the APEGT to the electric theory by putting $\rho_{\rm R}=0$, in which case we have
\begin{eqnarray}
 {\cal L}_A
 = -\frac{1}{4g_{\rm R}^2} \left[(\bar f_{\rm R})_{\mu\nu}^i\right]^2 ,
\end{eqnarray}
where we have rescaled the field by the coupling constant.
Hence we can avoid the unnatural argument used for deriving this form given in Eq.~(2.58) of Ref.~\citen{KondoI}.
We thus find that the APEGT is expected to be closed at each loop.
\par
In a similar way, we can switch the APEGT to another theory by setting $\rho_{\rm R}=1$, which yields
\begin{eqnarray}
 {\cal L}_B
 =     -\frac14\left[(\bar B_{\rm R})_{\mu\nu}^i\right]^2
    +\frac{i}{2g_R} (\bar B_{\rm R})_{\mu\nu}^i
     {}^\ast\!(\bar f_{\rm R})^{\mu\nu i} ,
\end{eqnarray}
without an unusual renormalization factor.
This is a candidate of the magnetic theory, as demonstrated in \S{}IV of Ref.~\citen{KondoI}.
Moreover, after decomposing the anti-symmetric tensor $B_{\mu\nu}^i$ into the (dual) gauge field $b_\mu^i$ using the Hodge decomposition and integrating out $b_\mu^i$,\footnote{To be more precise, we must integrate out the moduli characterizing the solution of the equation of motion when $b_\mu^i$ is regarded as a solution of the equation of motion.}
 we obtain the monopole action,
\begin{eqnarray}
  S_M [k] = \int d^4x \int d^4y
             {1 \over g_R^2} k^\mu(x) D_{\mu\nu}(x,y) k^\nu(y) ,
\end{eqnarray}
where $D_{\mu\nu}(x,y)$ is the massless vector propagator and $k^\mu$ is the magnetic monopole current defined by
\begin{equation}
  k^\mu := \partial_\nu {}^*\! \bar f_R^{\mu\nu} .
\end{equation}
This derivation should be compared with that of Eq.~(4.4) in Ref.~\citen{KondoI}.
\par
An off-diagonal gluon mass is generated by the quartic ghost interaction, which is necessary for renormalizability,\cite{MLP85} since the MA gauge is a nonlinear gauge (see Refs.~\citen{KS00a,Schaden} and \citen{HN93} for details).  It is desirable to include this effect in the APEGT within the above framework.
Such a study will be reported in Ref.~\citen{KS00c}.  Another important question to be answered is whether the above derived structure of the APEGT is preserved at the two-loop level.
These issues too will be discussed in Ref.~\citen{KS00c}.

\section*{Acknowledgements} %
The authors would like to thank a referee for bringing their attention to Ref.~\citen{BS88}.
This work is supported in part by a Grant-in-Aid for Scientific Research from the Ministry of
Education, Science and Culture (10640249).


\end{document}